\documentclass[aps,prl,twocolumn]{revtex4}
\usepackage{amssymb}
\usepackage{amsmath}
\usepackage{graphicx}
\usepackage{epsfig,amsmath}

\begin{document}

\title{Vortex Fluid Relaxation Model for Torsional Oscillation Responses of
Solid $^4$He }
\author{Sergey K. Nemirovskii$^{1,2}$, Nobutaka Shimizu$^{1}$, Yoshinori
Yasuta$^{1}$, and Minoru Kubota $^{1}$ }
\affiliation{$^{1}$Institute for Solid State Physics, University of Tokyo, Kashiwa,
Chiba277-8581, Japan \\
$^{2}$Institute of Thermophysics, Lavrentyev ave., 1, 630090, Novosibirsk,
Russia}
\date{\today }

\begin{abstract}
A phenomenological model is developed to explain 
new sets of detailed
torsional oscillator data for hcp $^4$He. The model is based on 
Anderson's idea of a vortex fluid(vortex tangle) 
in solid $^4$He. %
Utilizing a 
well-studied treatment of dynamics of quantized vortices we describe how the
"local superfluid component" 
is involved in rotation(torsion oscillations) via a polarized vortices
tangle. The polarization in the tangle appears both due to alignment of 
the remnant or thermal 
vortices and due to penetration of additional vortices into the volume. Both are
supposed to occur in a relaxation manner and the inverse full relaxation
time $\tau^{-1}$ is the sum of them. One of them is found to change linearly
with respect to the rim velocity $V_{ac}$. %
The developed approach explains the behavior of both $NLRS$ and $\Delta
Q^{-1}$ seen 
in the experiment. We 
reproduce not only the unique $V_{ac}$ dependence, but also obtain new
information about 
the vortices tangle, namely 
a divergence
in $\tau$ 
at extrapolated 
$T$$\sim $30 mK.
\end{abstract}

\maketitle

After the first report on \textquotedblright non-classical rotational
inertia\textquotedblright\ (NCRI) in solid $^{4}$He samples\cite{Kimchan 1},
the  confirmation came from several torsional oscillation(TO)
experiments, including some by the present authors \cite{list}. This finding
had been discussed in connection to the $NCRI$\cite{Leggett70} of a
supersolid as originally proposed by Leggett\cite{review}.
The measured drop of the period is expected to appear due to the reduction
of momentum of inertia, which originates 
in the appearance of
a superfluid component which does not follow the rotation of the sample
cell wall. 
It was, however, at problematically high $T$ where the "transition" was
reported. The following findings of the real onset temperature $T_{o}$\cite%
{VF} and possible vortex fluid(VF) state below $T_{o}$\cite{VF, Anderson-all}%
, 
would overcome the too high $T_{c}$ for BEC. And recent our work\cite{VFtoSS}
demonstrated a 
transition from the VF state to the supersolid state(SS) in solid $^{4}$He
below about 75 mK. In the present work we discuss the vortex dynamics in the
VF state, including the supercooled condition or measurements under
"equilibrium conditions"\cite{VFtoSS}. 
On the other hand, there have been 
various measurements by now showing that this phenomenon depends on various
parameters like 
pressure, impurity, sample quality\cite{upperlimit}, "orientation"\cite%
{orientation} etc. Dependence of the phenomenon on the excitation amplitude
of the TO (or on the rim velocity $V_{ac}$) 
gives rise to 
special interest as previously discussed
\cite{VF}. 
Fig.1 shows our TO results for the $V_{ac}$ dependence of both $NCRI$ (namely
relative change of period $\Delta P/P$) and dissipation $\Delta Q^{-1}$
(inverse quality factor) of the 49 bar sample, the former(lower column) was
given in our experimental work\cite{VF} and discussed as 
evidence for the VF state, pointing out the $log$$V_{ac}$ linear dependence,
originally proposed by Anderson\cite{Anderson-all}. In addition, the unique
feature that signal decreases when $V_{ac}$ is increased, is argued to be
evidence for 
thermally excited vortices in the VF state.
\begin{figure}[tbp]
\includegraphics[width=0.75
\linewidth]
{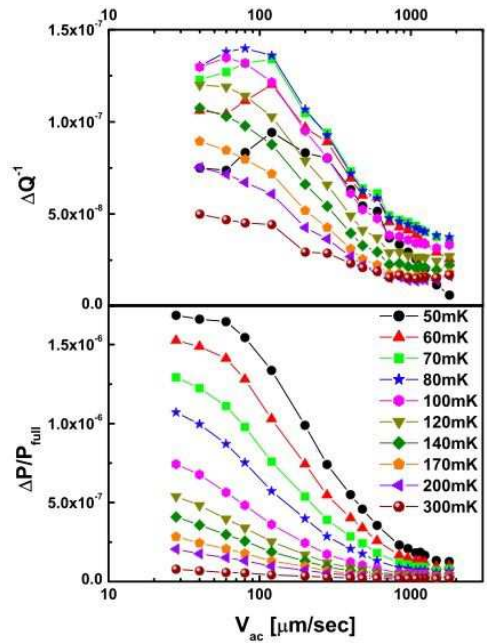}
\caption{New Data set of TO Responses throughout the vortex fluid state%
\protect\cite{VF}, including supercooled condition. Upper column indicate
energy dissipation$\Delta Q^{-1}$ and the lower column shows nonlinear
rotational susceptibility, $NLRS$=$\Delta $$P$/$\Delta $$P_{load}$ for 49
bar hcp $^{4}$He at different $T$'s as functions of $V_{ac}$.}
\label{Experiment}
\end{figure}
\indent  Let us discuss some properties of the observed phenomena\cite{VF} depicted
in Fig.1. \textcircled{1}.It is easy to see that the $V_{ac}$ dependence of the
period drop disappears for some characteristic velocities $V_{ac}$$\lesssim
10-30\ \mu m/s.$ \textcircled{2}.The drop of period (NLRS) decreases as the
applied $V_{ac}$ increases. This is a sign that the superfluid part is being
gradually involved in rotation. \textcircled{3}.For steady rotation with
velocities exceeding a characteristic 
velocity this effect vanishes, the sample rotates as a whole. \textcircled{4}%
.The characteristic value of the ratio $\frac{\Delta P}{P}/\Delta Q^{-1}$ is
$T$ and the pressure dependent quantity of order of unity. \textcircled{5}%
.At high $T$, 
the dissipation $\Delta Q^{-1}$ is a monotonically decreasing function of $%
V_{ac}$, whereas at low $T$ there is an obvious maximum. \textcircled{6}.One
more feature among reported results is the frequency dependence of both $P$
and 
$\Delta Q^{-1}$\cite{Aoki2007}. In the literature there was a speculation
that this behavior can be associated 
with quantized vortices. For instance, in \cite{KC science 2004} it is
pointed out 
that velocity $V_{ac}\approx 10\ \mu m/s$ coincides with the velocity
created by a single circulation around the sample.
Prokof'ev\cite{review} pointed out that \ "To understand why NLRS decreases
with $V_{ac}$, one has to consider the non-linear response of vortex loops
and pinned vortex lines to the flow". Huse et al.\cite{Huse2007} developed a
simple phenomenological model which introduces dissipative relative motion
of two components realized via \textquotedblleft phase
slips\textquotedblright\ of quantized vortices. P.W. Anderson\cite%
{Anderson-all} describes the scenario of 
a set of chaotic vortices (vortex fluid or vortex tangle) which under the
torsional oscillation(TO) 
behaves like vortex-anivortex pairs in the Kosterlitz-Thouless model, and
free (unbalanced) vortices bring the 
superfluid part into rotation. The role of vortices in rotation and
torsional oscillation of solid helium had been discussed also in \cite%
{saslow04,pomeau}.\\
\indent    In the following, we propose a phenomenological model describing the
behavior of the torsional oscillations 
in the presence of a vortex tangle. In a vortex free sample or in the case
of an absolutely isotropic vortex tangle, the superfluid fraction does not
participate in rotation or torsional oscillations.
Therefore the momentum of inertia $I_{full}$ acquires a deficit $%
I_{SF}=\rho _{s}VR^{2}/2$ where $\rho _{s}$ is the 'local' superfluid density for the VF state, and $V
$ is the volume of the sample. Angular momentum of the superfluid fraction
appears only due to the presence of either aligned vortices (vortex array)
or due to the polarized vortex tangle having nonzero total average
polarization $\mathbf{P=}\mathcal{L}\left\langle \mathbf{s}_{z}^{\prime
}(\xi )\right\rangle $ along the applied angular velocity $\mathbf{\Omega }$
(axis $z,$ the magnitude of ${\Omega }=V_{ac}/R$, where R is radius of the
sample). Here $\mathcal{L}$ is the vortex line density (total length per
unit volume), $\mathbf{s}(\xi )$ is the vector line position as a function
of label variable $\xi $, $\mathbf{s}^{\prime }(\xi )$ is the tangent
vector. In the \emph{steady case} there is a strictly fixed relation between
the total polarization $\mathcal{L}\left\langle \mathbf{s}^{\prime }(\xi
)\right\rangle $ and applied angular velocity $\mathbf{\Omega }$,
\begin{equation}
\mathbf{\Omega }=\kappa \mathbf{P/}2=\kappa \mathcal{L}\left\langle \mathbf{s%
}^{\prime }(\xi )\right\rangle /2.  \label{polarization}
\end{equation}%
Here $\kappa $ is the quantum of circulation. In a case when vortex
filaments form an array the quantity $\mathcal{L}$ coincides with $2D$
density $n$ and (\ref{polarization}) transforms to the usual Feynman's
rule.\ Angular momentum of the superfluid part can be written as $\mathbf{M}%
_{SF}\mathbf{=}I_{SF}\mathbf{\Omega =}I_{SF}\kappa \mathbf{P/}2$.\\
\indent    The situation drastically changes in a nonstationary (transient or
oscillating) case. The total polarization $\mathbf{P}(t)$ changes in time
owing to 
both the vortex line density $\mathcal{L(}t)$ and the mean local \
polarization $\left\langle \mathbf{s}^{\prime }(t)\right\rangle $ change in
time according to their own, relaxation-like dynamics. Therefore the angular
momentum of the superfluid part is not $\mathbf{M}_{SF}\mathbf{=}I_{SF}%
\mathbf{\Omega }$ anymore. Because of relaxation processes there is
retardation between $\mathbf{\Omega }(t)$ and $\mathbf{M}_{SF}\mathbf{(}t)$,
\ and the connection between them is nonlocal in time and $\mathbf{M}_{SF}%
\mathbf{(}t)$ is some functional of time dependent angular velocity $%
\mathbf{\Omega }(t)$. There are two possible mechanisms for relaxation-like
polarization of the vortex fluid. The first is an alignment of elements of
the vortex 
lines due to interaction with the normal component (See \cite{Tsubota
combination} for detailed explanations). This interaction (mutual friction)
is proportional to the local normal velocity , which in turn is proportional
to the rim velocity $\mathbf{V}_{ac}$. Thus it is natural to suppose that
polarization $\mathbf{P}$ of the vortex tangle due to alignment of filaments
along $\mathbf{\Omega }(t)$ occurs with typical inverse time $\tau _{1}^{-1}(%
\mathbf{V}_{ac})$ which is proportional to the rim velocity $\mathbf{V}_{ac}
$. Let us illustrate the  above with consideration performed in \cite%
{Tsubota combination}. In the presence of mutual friction there is a torque
acting on the line and the angle $\phi $ between axis $z$ and the line
element changes according to the equation$\ d\phi /dt=\alpha (\mathbf{V}%
_{ac}/R)\sin \phi $ ($\alpha $ is the friction coefficient, dependent, in
general, on 
$T$\ and pressure $p$). Except for a short transient, 
the solution to this equation can be described as a pure exponential $\ \sim
\exp (-t/\tau _{1}(\mathbf{V}_{ac}))$, with the velocity dependent inverse
time $\tau _{1}^{-1}(\mathbf{V}_{ac})\sim \alpha \mathbf{V}_{ac}/R$. Thus,
we conclude that during time-varying 
rotation or torsional oscillation vortex filaments tend to
align along the angular velocity direction. However, there can be not enough
pre-existing vortex lines in the tangle to involve all the superfluid part
into the rotation to satisfy the relation (\ref{polarization}), or on the
contrary the initial vortex tangle can be excessively dense. In this case
deficient (extra)) vortices should penetrate into (leave from) the bulk of the sample. This
penetration occurs in a 
diffusion-like manner\cite{Nem_diffusion}\ and leads to the relaxation-like
saturation of the vortex line density $\mathcal{L}(t)$ \ We assume 
that this saturation occurs in an exponential manner with some
characteristic inverse time $\tau _{2}^{-1}=\beta $. Due to linearity of the
diffusion process we suppose that coefficient $\beta $ is velocity
independent, but can be a function of $T$\ and $p$. Combining 
both mechanisms we assume that \ the whole polarization of the vortex fluid
occurs in the relaxation manner with pure exponential behavior 
$\varphi (t^{\prime }/\tau )\sim \exp (t^{\prime }/\tau )$, and the inverse
time $\tau ^{-1}$ of relaxation is just the sum of $\tau _{1}^{-1}({V}_{ac})$
and $\tau _{2}^{-1}$,
\begin{equation}
\tau ^{-1}=\alpha (T)\mathbf{V}_{ac}/R+\beta (T)  \label{inverse time}
\end{equation}%
\indent    In the presence of relaxation the angular momentum $\mathbf{M(}t)$ of the
superfluid part is related to the applied angular velocity $\mathbf{\Omega }%
(t)$ by the nonlocal relation,
\begin{equation}
\mathbf{M}=a\mathbf{\Omega }(t)+b\int\limits_{0}^{\infty }\mathbf{\Omega }%
(t-t^{\prime })\varphi (\frac{t^{\prime }}{\tau })\frac{dt^{\prime }}{\tau }.
\label{causality}
\end{equation}%
Relation (\ref{causality}) implies that the angular momentum $\mathbf{M(}t)$
depends on the applied angular velocity $\mathbf{\Omega }(t)$ taken in the
all previous moments of time with the weight $\exp (-t/\tau )$. To clarify
the physical meaning 
of constants $a$ and $\ b$ we consider the limiting cases of very small and
very large frequencies. In case $\omega \rightarrow 0$ 
the slowly changing function $\mathbf{\Omega }(t-t^{\prime })$ can be
considered as a constant and be taken out of the integral, whereupon the
rest of integral becomes 
unity and we have $\mathbf{M}_{\omega \rightarrow 0}=(a+b)\mathbf{\Omega }$.
But at the same time, 
both components participate in the solid body rotation, thus $(a+b)=I_{full}$.
In the opposite case of very large frequencies, $\omega $$\rightarrow $$%
\infty $, the integral from rapidly oscillating functions $\mathbf{\Omega }%
(t-t^{\prime })$ vanishes, so $\mathbf{M}_{\omega \rightarrow \infty }=a%
\mathbf{\Omega }$. \ Since under these conditions the superfluid component
does not participate in the motion at all, we conclude that the constant $a$
is nothing but the full moment of inertia\ $I_{N}$ of the sample without the
superfluid part (which includes momentum of inertia of the empty cell $%
I_{empty}$ ). Thus, the quantity $b$ is moment of inertia\ $I_{SF}$ of the
superfluid part. \ Substituting (\ref{causality})\ with $a$ $=\ I_{N}$ and $b
$ $=\ I_{SF}$ into the equation of motion of the TO, we get%
\begin{equation}
\frac{d}{dt}\left[ I_{N}\mathbf{\Omega }(t)+I_{SF}\int\limits_{0}^{\infty }%
\mathbf{\Omega }(t-t^{\prime })\varphi (\frac{t^{\prime }}{\tau })\frac{%
dt^{\prime }}{\tau }\right] +k\theta =0.  \label{equation of motion}
\end{equation}%
Here $\theta (t)$ is the angle of rotation of the oscillator, $k$ is the
spring constant. Relation(\ref{equation of motion}) is an
integro-differential equation and, in general, not easy to solve.
Because $\varphi (\frac{t^{\prime }}{\tau })$ is a pure exponential function
we can eliminate 
the integral term. Omitting details we arrive at 
the case where  equation (\ref{equation of motion}) is reduced to an
ordinary differential equation of the third order, which has a solution in
the form $\theta (t)=\theta _{0}\exp (i\omega t)$. The frequency $\omega $
satisfies the relation%
\begin{equation*}
\omega =\sqrt{\frac{k}{I_{full}}}(1+\frac{I_{SF}}{2I_{full}}\frac{\left(
\omega \tau \right) ^{2}}{(\omega \tau )^{2}+1)}+\frac{I_{SF}}{2I_{full}}%
\frac{i\omega \tau }{(\omega \tau )^{2}+1)}).
\end{equation*}%
\indent    Thus, the frequency of the oscillation consists of three parts. The first
one $\omega _{0}=\sqrt{k/I_{full}}$ describes the oscillation with full
moment of the inertia $I_{full}$ as if all ingredients (empty cell, normal
part, superfluid part) fully participate in motion. The second term is
responsible for 
increase of the frequency because the superfluid component  participates
in the torsional oscillation only partly. The third term is the imaginary one.
It describes the attenuation of the oscillation amplitude, i.e. it describes
the dissipation. The amplitude decreases (with time) as $\exp \left[ -\Im
(\omega )t\right] ,$ and the inverse quality factor is $Q^{-1}=\frac{2\Im
(\omega )}{\omega }$. Using the smallness of the $I_{SF}<<I_{full}$ we put $%
\omega =\omega _{full}$ in the right hand side, yielding (index in $\omega
_{full}$ is omitted)
\begin{equation}
\frac{\Delta P}{P}=-\frac{1}{2}\frac{I_{SF}}{I_{full}}\frac{\left( \omega
\tau \right) ^{2}}{(\omega \tau )^{2}+1}.  \label{period drop}
\end{equation}%
\begin{equation}
\Delta Q^{-1}=\frac{2\Im (\omega )}{\omega }=\frac{I_{SF}}{I_{full}}\frac{%
(\tau \omega )}{\tau ^{2}\omega ^{2}+1}\ \ \   \label{inverse quality}
\end{equation}%
\indent     Relations (\ref{period drop}),(\ref{inverse quality}) are the final solution
to the problem of the torsional oscillation when the superfluid component is
involved in rotation via polarized vortex fluids, and polarization occurs in
the relaxation-like manner. Being phenomenological, the approach developed
does not allow determining some quantities entering the formalism. Thus the
parameters $\alpha (T,p)$\textbf{\ }and\textbf{\ }$\beta (T,p)$ responsible
for the relaxation of the vortex tangle 
should be also obtained on the basis of the approach describing dynamics of
quantized vortices, which is so far absent. Nevertheless comparison of our
results with the experimental data allows us to explain a series of
experimental results and to get some quantitative information and insights.
Let us analyze relations (\ref{period drop}) and (\ref{inverse quality}).
From relations (\ref{period drop}),(\ref{inverse quality}) of our paper it
follows that $\frac{\Delta P}{P}/\Delta Q^{-1}$ is equal to $(1/2)(\omega
\tau )$. It can take any value depending on the arrangement of the experiment.
But usually observations should be under conditions with $\omega
\tau $ on the order of unity. Therefore in many experiments the $\Delta P/P/$
and $\Delta Q^{-1}$ are of the same order of magnitude, although sometimes they
can be significantly different (see \cite{Huse2007}). Dividing the first
relation by the second one and taking the zero $\mathbf{V}_{ac}$ limit \ in the
relation $\frac{\Delta P}{P}/\Delta Q^{-1}$ = $(1/2)(\omega \tau )$ we get
an expression for relaxation time $\beta (T)$ due to diffusion of vortices.
\begin{figure}[tbp]
\includegraphics[width=1.0
\linewidth]
{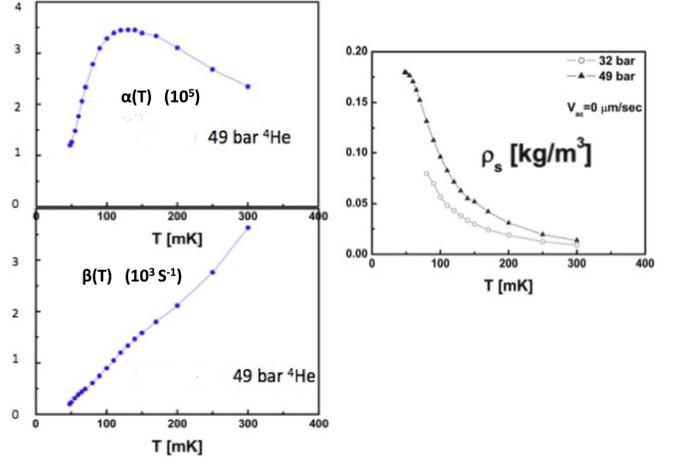} \label{rhos}
\caption{Parameters $\protect\alpha (T)$, $\protect\beta (T)$, and $\protect%
\rho _{s}(T)$ obtained from the data of Fig.1 with the use of analysis
described in text. $\protect\beta (T)$ goes to zero, or $\protect\tau $ to
infinity at extrapolated $T$$\approxeq 30$ mK.}
\end{figure}
Taking further the zero $\mathbf{V}_{ac}$ limit for the period drop, and
assuming that $\beta (T)$ abruptly vanishes below the 'critical velocity'
(which is equivalent to absence of vortices), we find the superfluid
momentum of inertia $I_{SF}$, and, consequently superfluid density $\rho _{s}
$ can be extracted from the graphs for $\Delta P/P$. Knowing $I_{SF}$ $(\rho
_{s}(T))$, $\beta (T)$ and fitting the curves $\Delta P/P$ as functions of $%
\mathbf{V}_{ac}$ it is possible to determine the inverse relaxation time due
to aligning $\ \tau _{1}^{-1}(\mathbf{V}_{ac})\sim \alpha (T)\mathbf{V}%
_{ac}/R$ and quantity $\alpha (T)$. Performing all procedures described
above, we have all the necessary data,
as shown in Fig. 2 where parameters $t\alpha (T)$, $\beta (T)$, and $\rho
_{s}(T)$ are depicted.\\
\indent   In Fig. 3 we show $\Delta Q^{-1}$ and $\frac{\Delta P}{P}$=$NLRS$ as
functions of 
$\mathbf{V}_{ac}$, drawn 
using relations (\ref{period drop}) and (\ref{inverse quality}) and
extracted experimental data. It can be seen that shapes of curves and their
response to the change of $T$ 
correspond to the curves shown in Fig.1 and Fig.2. It is seen that in the
limit $\mathbf{V}_{ac}\rightarrow 0,$ or $\omega \rightarrow \infty $, or $%
\alpha (T)\rightarrow 0$, $NLRS$ 
reaches 
the maximum value. Physically it is clear, since under these conditions the
superfluid part cannot 
participate in rotation at all. \ Other limits $\mathbf{V}_{ac}\rightarrow
\infty ,$ or $\omega \rightarrow 0$, or $\alpha (T)\rightarrow \infty $
correspond to the vanishing of the effect, which is also reasonable since
under these conditions the superfluid part participates in the solid body
rotation, and 
no effect appears.
\begin{figure}[tbp]
\includegraphics[width=0.75
\linewidth]
{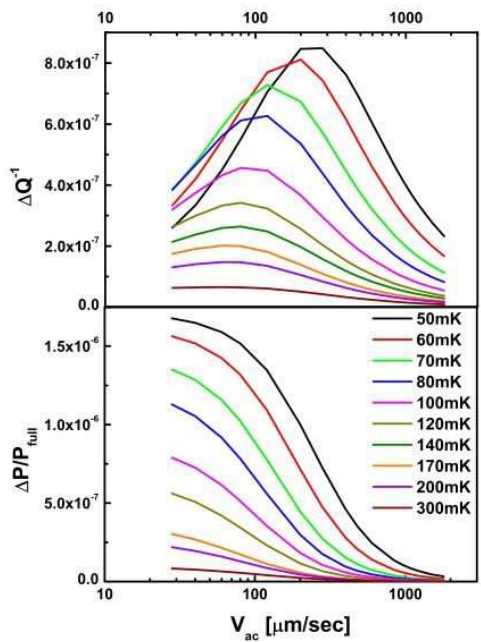} \label{Calc}
\caption{Energy dissipation$\Delta Q^{-1}$and nonlinear rotational
susceptibility NLRS at different $T$ as a function of $V_{ac}$, obtained
using relations (\protect\ref{period drop}),(\protect\ref{inverse quality})
with parameters taken from Fig. 2}
\end{figure}
If relaxation due to diffusion (penetration) is weak, then the dependence of
dissipation becomes non-monotonic. Analysis shows that the critical value of
$\tau _{2}^{-1}=\beta (T)$ \ is equal to the frequency $\omega $. In fact,
this can differ by some factor on the order of unity. One of the possible
reasons for this difference is that we calculate using a purely exponential
relaxation process, whereas in reality it can be better described by a more complicated 
dependence. The maximum value of dissipation $\Delta $$Q_{peak}^{-1}$ should
be at $\frac{1}{2}$ $\frac{\Delta P}{P}$ and it should be reached at values
of the rim velocity $\mathbf{V}_{ac}=R(\omega -\beta (T))/\alpha (T)$. This
tendency is easily seen in 
Fig. 1, $\Delta $$Q_{peak}^{-1}$ decreases with
$T$ and shifts in the direction of small $V_{ac}$, 
then for some "critical temperature" when $\tau _{2}^{-1}=\beta (T)$, the
peak disappears entirely. It happens at $T$ about $120\ mK$. Comparing with
the experimental data one can conclude that the behavior described above
indeed takes place for $T$ above about $75\ mK$, but the agreement fails for
lower $T$. It is remarkable that $75\ mK$ 
was detected by 
authors of the present paper, as $T_{c}$ 
below which a hysteretic behavior takes place as a sign of
a transition to a supersolid(SS) state(see \cite{VFtoSS}). Relations (\ref%
{period drop}) and (\ref{inverse quality}) can also explain the $f$ = $%
\omega $/2$\pi $ 
dependence of $NLRS$ and $\Delta $$Q^{-1}$ observed in \cite{Aoki2007}.
Indeed, the significant dependence on $\omega $ appears when the inverse
time ~$\tau ^{-1}$ of relaxation is comparable with $\omega $, which can
happen at higher $T$. In this range 
of parameters, the 
$\Delta $$P/P$ (\ref{period drop}) 
is a monotonic function of $\omega $.\newline
In summary the phenomenological model of relaxation processes of the VF
state has been introduced. Unsteady rotation and torsional oscillation
have been studied. Dependence of both the $NLRS$
and the $\Delta $$Q^{-1}$ 
on $T$, $V_{ac}$ and $f$ 
have been studied. The results obtained may serve as a good qualitative
description of the corresponding 
measurements in the VF 
state in solid $^{4}$He. Combining theoretical predictions with experimental
data it became possible to obtain some quantitative results. Actually recent
experimental results\cite{superglass} can be well understood in terms of the
present VF 
analysis, as an alternative to the 
interpretation in terms of superglass, by other authors.\\
\indent     The authors acknowledge A. Penzev for his contributions in the early stage
and R.M. Mueller for help. S.N. thanks the Institute for Solid State
Physics(ISSP), Univ. of Tokyo for 
the ISSP visiting Professor program. This work was partially supported by
grant 07-02-01124 from the RFBR and grant of scientific schools 4366.2008.8

\end{document}